\documentclass[11pt]{article}
\usepackage{moriond,epsfig}

\def\Journal#1#2#3#4{{#1} {\bf #2}, #3 (#4)}

\def\AP{\em Annals of Phys.}
\def\JHEP{\em Jour. of High Energy Phys.}

\def\NPB{{\em Nucl. Phys.} B}
\def\PA{{\em Physica A}}
\def\PLB{{\em Phys. Lett.}  B}
\def\PRL{\em Phys. Rev. Lett.}
\def\PRD{{\em Phys. Rev.} D}
\def\ZPC{{\em Z. Phys.} C}

\def\be{\begin{equation}}
\def\ee{\end{equation}}
\def\bea{\begin{eqnarray}}
\def\eea{\end{eqnarray}}
\def\p{\pi}
\def\t{\tau}
\def\m{\mu}
\def\n{\nu}

\newcommand{\ket}{\,\rangle}
\newcommand{\bra}{\langle \,}
\begin{document}
\vspace*{4cm}
\title{HADRONIZATION IN $\t\to KK \pi\n_\t$ DECAYS}

\author{ P. ROIG }

\address{Instituto de F\'isica Corpuscular, IFIC, CSIC-Universitat de Val\`encia.\\ Apt. de Correus 22085, E-46071 Val\`encia, Spain}

%%%%%%%%% authors photo (option) %%%%%%%%%%%%%%%%%%%%% 
%\begin{figure}[h]
%\begin{center}
%\includegraphics[scale=0.65]{Pablo.eps}
%\framebox[35mm]{\rule[-11mm]{0mm}{35mm}}
%\end{center}
%\end{figure}
%%%%%%%%%%%%%%%%%%%%%%%%%%%%%%%%%%%%%%%%%%%%%%%%%%%%%

\maketitle\abstracts{
Hadronization in $\t\to KK \pi\n_\t$ decays is driven by both vector and axial-vector currents that we study, guided by the following principles: The 1/$N_C$ expansion -worked out at leading order, considering only the contribution of the lightest spin one resonances-, approximate chiral symmetry at low energies and the appropriate asymptotic behaviour we demand to the associated form factors. All these features are implemented in the resonance theory. Most of its couplings are determined by imposing the short-distance requirements of vector and axial-vector spectral functions within QCD. We plan to improve our prediction of the hadronic spectra using recently available experimental data.}

\section{Introduction}
The mass of the $\t$ lepton, $\sim1.8$ GeV, makes it privileged for it connects the hadronic and leptonic worlds through its semileptonic decays in the cleanest possible way. On the other hand, it sets the typical energy scale of these processes in a region where the fundamental theory of strong interactions, QCD~\cite{qcd}, becomes computationally useless.\\
One may, of course, rely on parameterizations of the assumed dynamics that have proved to be very successful in the modes involving two and three pions, namely the K\"uhn-Santamar\'i{}a (KS) model \cite{ks}, but it would be desirable to explain the experimental data within a framework that resembles QCD as much as possible.\\
On the very low energy domain $\chi$PT \cite{wei}$^,$ \cite{cpt} is the QCD dual at energies well below the $\rho$(770) mass, i.e., $E\ll0.6$ GeV. As expected, however, $\chi$PT only applies to a small part of the phase space in $\t$ decays \cite{col}. The involved hadronization can not be understood without including explicitly the resonances that are exchanged in the process. The most common procedure is to rely on the phenomenologically successful notion of vector meson dominance \cite{vmd}, as we do.\\
The easiest way to incorporate the resonances is through Breit-Wigner functions, including phase-space motivated off-shell widths and a weighted interplay between the fundamental and excited resonances as in Ref.~\cite{ks}. This has been shown \cite{p03}$^,$ \cite{t3p} to violate chiral symmetry at NLO in the chiral expansion. In addition, the assumed structure of exchanged resonances in the $KK\pi$ modes \cite{fm} does not cover all those allowed, as can be checked by comparison with Ref.~\cite{paper}. Furthermore, Ref.~\cite{fm} includes different mass and width for the $\rho$(770) depending on its appearance on the axial-vector current or the vector one. Finally, two multiplets of vector resonances are used \cite{fm} in the axial-vector current, while three of them are employed in the vector one. The aim of our work is to achieve a description of the considered decays closer to QCD than this one.\\
CLEO \cite{cleo}, BaBar \cite{babar1} and Belle \cite{belle} have been collecting good quality data on $\t\to KK \pi\n_\t$ decays. CLEO \cite{cleo} announced that the parameterization in Ref.~\cite{fm} was unable to fit their data so that they reshaped it violating the Wess-Zumino normalization emanating from the chiral anomaly of QCD, as was put forward in Ref. \cite{p04}. BaBar has not published these results yet. However, they have managed \cite{babar2} to split with great precission the isoscalar and isovector contributions in $e^+e^-\to KK\pi$ so that, under the isospin symmetry we can use it for the considered $\t$ decays, as it has been done in Ref.~\cite{mal}.\\
The fact that CLEO and BaBar disagree with respect to their respective results being explained or not by a model generalizing KS~\cite{ks}, on the one hand, and the inconsistencies in the treatment followed in Refs. \cite{fm}$^,$\cite{fm2} -as stated in the previous paragraph- are the motivations for our work trying to explain the experimental measurements while learning as much as possible about the hadronization of the involved QCD currents.\\

\section{THE RESONANCE CHIRAL THEORY OF QCD} \label{RChT}
As opposed to the very low energy sector, the effective theory of QCD at intermediate energies remains unknown. Our aim is to resemble the underlying theory as much as possible. For this, we rely on Resonance Chiral Theory (R$\chi$T), as introduced in the article \cite{rcht1}, which is built upon the approximate chiral symmetry of low-energy QCD for the lightest pseudoscalar mesons and unitary symmetry for the resonances. Weinberg's Theorem \cite{wei}$^,$ \cite{leut} ensures the correctness of writing a phenomenological Lagrangian in terms of mesons fulfilling the basic symmetries of light-flavoured QCD. Its large-$N_C$ limit \cite{lnc} guides the perturbative expansion of R$\chi$T. At LO in 1/$N_C$, meson dynamics is described by tree level diagrams obtained from an effective local Lagrangian including the interactions among an infinite number of stable resonances. In $\t$ decays we need to include finite resonance widths (a NLO effect in this expansion) as we do within our framework \cite{width}. We are also departing from the $N_C\to\infty$ limit because we consider just one multiplet of resonances per set of quantum numbers (single resonance approximation \cite{sra}) and not the infinite tower predicted that cannot be included in a model independent way. Representing the resonances in the antisymmetric tensor formalism proves to be more convenient \cite{rcht2} in processes involving mesons as asymptotic states, instead of the Proca \cite{vec} or mixed \cite{mix} ones.\\
The relevant part of the R$\chi$T Lagrangian is \cite{rcht1}$^,$ \cite{vap}$^,$ \cite{vvp}$^{,}$ \cite{tesina}:
\bea \label{Full_Lagrangian}
\mathcal{L}_{R\chi T}=\frac{F^2}{4}\bra u_\mu u^\mu +\chi_+ \ket+\frac{F_V}{2\sqrt{2}}\bra V_{\mu\nu} f^{\mu\nu}_+ \ket +
\frac{i G_V}{\sqrt{2}} \bra V_{\mu\nu} u^\mu u^\nu\ket + \frac{F_A}{2\sqrt{2}}\bra A_{\mu\nu} f^{\mu\nu}_- \ket \nonumber\\
+\mathcal{L}_{\mathrm{kin}}^V+\mathcal{L}_{\mathrm{kin}}^A + \sum_{i=1}^{5}\lambda_i\mathcal{O}^i_{VAP} + \sum_{i=1}^7\frac{c_i}{M_V}\mathcal{O}_{VJP}^i +\sum_{i=1}^4d_i\mathcal{O}_{VVP}^i + \sum_{i=1}^5  \frac{g_i}{M_V} {\mathcal O}^i_{VPPP} \, ,
\eea
where all couplings are real, being $F$ the pion decay constant in the chiral limit. The notation is that of Ref.~\cite{rcht1}. Here and in the following $P$ stands for the lightest pseudoscalar mesons and $A$ and $V$ for the (axial-)vector mesons. Furthermore, all couplings in the second line are defined to be dimensionless. For the explicit form of the operators in the last line, see \cite{vap}$^,$ \cite{vvp}$^,$ \cite{tesina}.\\
In order to inherit the maximum possible features of QCD, to be implemented in R$\chi$T, we still have to exploit the matching between order parameters of spontaneous chiral symmetry breaking with partonic QCD related quantities. The matching of $n$-point Green Functions in the OPE of QCD and in R$\chi$T has been shown to be a fruitful procedure \cite{rcht2}$^,$ \cite{vap}$^,$ \cite{vvp}$^{,}$ \cite{amo}$^{,}$ \cite{knecht}$^{,}$ \cite{consistent}. Additionally, we will demand to the vector and axial-vector form factors a Brodsky-Lepage-like behaviour \cite{brodskylepage}.\\
While symmetry fully determines the structure of the operators, it is the QCD-ruled short-distance behaviour who restricts certain combinations of couplings rendering R$\chi$T predictive provided we fix these few remaining parameters restoring to phenomenology: $F_V$ could be extracted from the measured $\Gamma(\rho^0\to e^+e^-)$, $G_V$ from $\Gamma(\rho^0\to \pi^+\pi^-)$, $F_A$ from $\Gamma(a_1\to \pi\gamma)$ and the $\lambda_i$'s from $\Gamma(a_1\to \rho\pi)$ which stars on the $\t\to3\pi\n_\t$ processes themselves. $\Gamma(\omega\to \pi\gamma)$, $\Gamma(\omega\to3\pi)$ and the $\mathcal{O}(p^6)$ correction to $\Gamma(\pi \to \gamma\gamma)$ may give us information on the remaining couplings \cite{vvp}.
\section{FORM FACTORS IN $\tau^-\to(KK\pi)^-\,\nu_\tau$} \label{FormFactors}
The decay amplitudes for all charge channels can be written as
\begin{equation} \label{Mgraltau}
\mathcal{M}\,=\,-\frac{G_F}{\sqrt{2}}\,V_{ud}\overline{u}_{\nu_\tau}\gamma^\mu(1-\gamma_5)u_\tau \mathcal{H}_\mu\,,
\end{equation}
where
\begin{eqnarray} \label{Hadronic_matrix_element}
\mathcal{H}_\mu & = & \bra P(p_1) P(p_2) P(p_3)|\left( \mathcal{V}_\mu - \mathcal{A}_\mu\right)  e^{i\mathcal{L}_{QCD}}|0\ket=\\
& & V_{1\mu} F_1^A(Q^2,s_1,s_2) + V_{2\mu} F_2^A(Q^2,s_1,s_2) + Q_\mu F_3^A(Q^2,s_1,s_2) + i V_{3\mu} F_4^V(Q^2,s_1,s_2)\,,\nonumber
\end{eqnarray}
and
\begin{eqnarray}
V_{1\mu}\, = \, \left( g_{\mu\nu} - \frac{Q_{\mu}Q_{\nu}}{Q^2}\right) \,
(p_2 - p_1)^{\nu} \;\;\;\; , \;\;\;\; V_{2\mu}\, = \, \left( g_{\mu\nu} - \frac{Q_{\mu}Q_{\nu}}{Q^2}\right) \,
(p_3 - p_1)^{\nu}\,,\nonumber\\
V_{3\mu} = \varepsilon_{\mu\nu\varrho\sigma}p_1^\nu\, p_2^\varrho\, p_3^{\sigma}\;\; \;\; , \;\;\;\; Q_\mu\,=\,(p_1\,+\,p_2\,+\,p_3)_\mu\;\;\;\;,\;\;\;\;s_i = (Q-p_i)^2\,.
\end{eqnarray}
Here $F_i$, $i=1,2,3$ correspond to the axial-vector current while $F_4$ drives the vector current. The form factors $F_1$ and $F_2$ have a transverse structure in the total hadron momenta, $Q^\mu$, and drive a $J^P=1^+$ transition. The scalar form factor, $F_3$, vanishes with the mass of the Goldstone bosons (chiral limit) and, accordingly, gives a tiny contribution. In Ref.~\cite{t3p} the vector form factor was not contributing, so we do not only intend to confirm or refuse the bounds on the $\lbrace \lambda_i\rbrace_{i=1}^5$, but to explore the vector current sector of the resonance Lagrangian. The explicit expressions of the form factors can be found in Refs.~\cite{tesina}$^,$ \cite{paper}.\\
\section{ASYMPTOTIC BEHAVIOUR AND QCD CONSTRAINTS} \label{Asymptotic_behaviour}
\begin{figure}[h]
\centering
\includegraphics[scale=0.65,angle=-90]{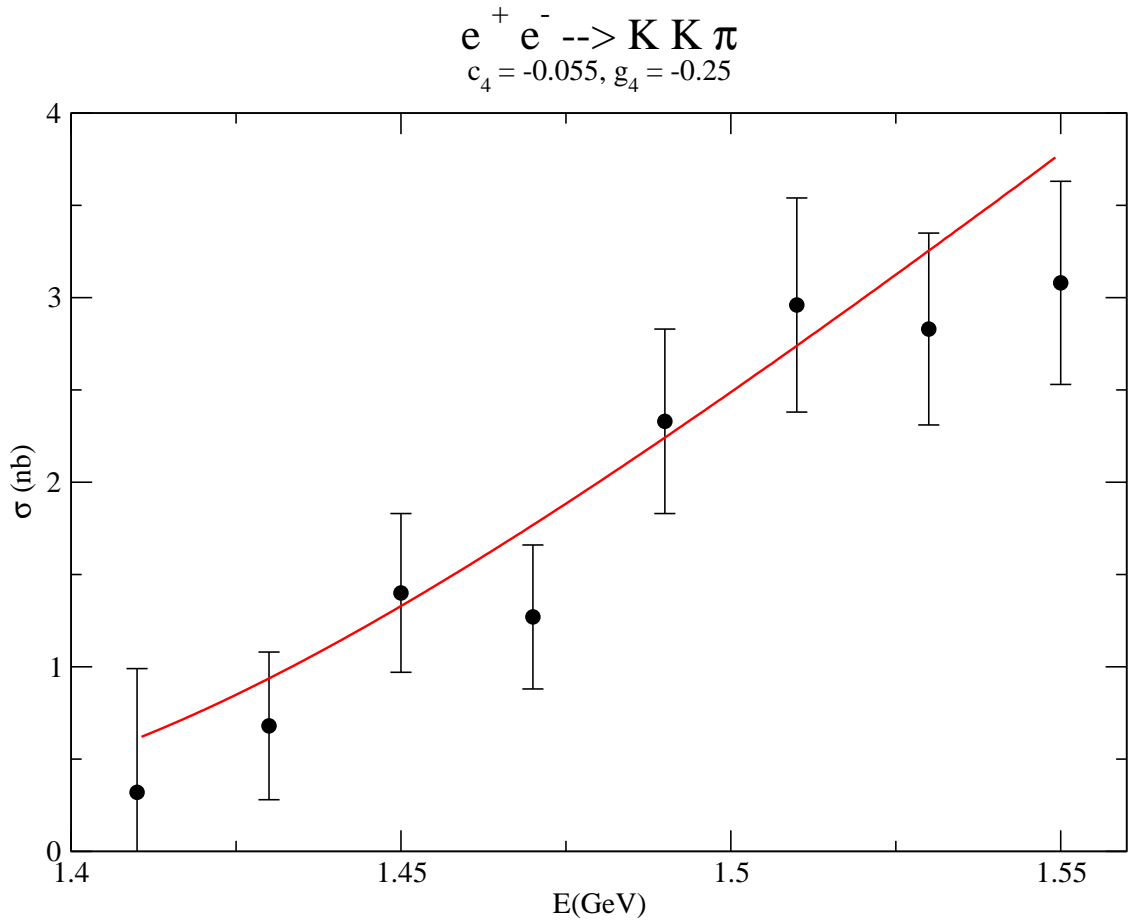}
\caption{\small{Fit of our computation for $\sigma\left( e^+e^-\to KK \pi\right)$ to BaBar data. At higher energies than those shown the contribution of excited resonances we do not have taken into account becomes relevant.}}
\label{fig}
\end{figure}

Computing the Feynman diagrams involved shows that the result depends only on three combinations of the $\lbrace \lambda_i\rbrace_{i=1}^5$, four of the $\lbrace c_i\rbrace_{i=1}^7$, two of the $\lbrace d_i\rbrace_{i=1}^4$ and four of the $\lbrace g_i\rbrace_{i=1}^5$. The number of free parameters has been reduced from 24 to 15.\\
\indent We require the form factors of the $A^\m$ and $V^\m$ currents into $KK\p$ modes vanish at infinite transfer of momentum. As a result, we obtain constraints \cite{proc} among all axial-current couplings but $\lambda_0$, that are also the most general ones satisfying the demanded asymptotic behaviour in $\t\to3\p\n_\t$. Proceeding analogously with the vector current form factor results in five additional restrictions\cite{proc}. From the 24 initially free couplings in Eq. (\ref{Full_Lagrangian}), only five remain free: $c_4$, $c_1\,+\,c_2\,+\,8\,c_3\,-\,c_5$, $d_1\,+\,8\,d_2\,-\,d_3$, $g_4$ and $g_5$. After fitting $\Gamma(\omega\to3\pi)$ -using some relations obtained in \cite{vvp}- only $c_4$ and $g_4$ remain unknown.\\
\indent Employing isospin symmetry \cite{aleph} we are able to provide a theoretical expression for $\sigma\left( e^+e^-\to KK \pi\right)$ that we have fitted to BaBar data \cite{babar2} obtaining $c_4=-0.052\pm 0.003$ and $g_4=-0.20^{+0.08}_{-0.12}$. Remarkably, this procedure has allowed us to be sensitive to the sign of $c_4$ improving the work done in \cite{proc}. The results are represented in Fig \ref{fig}.\\
\section*{Acknowledgments}
 I wish to thank Jean Tran Thanh Van for the excellent organization, and the warm and cosy atmosphere of the QCD and High Energy Interactions 2008 meeting in La Thuile (Italy). I acknowledge B.~Malaescu for pointing out to me that BaBar $e^+e^-\to KK\pi$ data was already available. This work has been done in collaboration with D. G\'omez-Dumm, A. Pich and J. Portol\'es. I wish to thank J.~P. for a careful revision of the manuscript. P.~R. is supported by a FPU contract (MEC). This work has been supported in part by the EU MRTN-CT-2006-035482 (FLAVIAnet), by MEC (Spain) under grant FPA2007-60323, by the Spanish Consolider-Ingenio 2010 Program CPAN (CSD2007-00042) and by Generalitat Valenciana under grant GVACOMP2007-156.\\

\end{document}